\documentclass[apj]{emulateapj}
\usepackage{subfigure}
\usepackage{graphics}

\def\spose#1{\hbox to 0pt{#1\hss}}
\def\simlt{\mathrel{\spose{\lower 3pt\hbox{$\mathchar"218$}}
    \raise 2.0pt\hbox{$\mathchar"13C$}}}
\def\simgt{\mathrel{\spose{\lower 3pt\hbox{$\mathchar"218$}}
    \raise 2.0pt\hbox{$\mathchar"13E$}}}

\newcommand{\oiiiw}{\mbox{[\ion{O}{3}] $\lambda$5007} $\,$}
\newcommand{\oiiiwn}{\mbox{[\ion{O}{3}] $\lambda$5007}}

\newcommand{\hb}{\mbox{H$\beta$} $\,$}
\newcommand{\hbn}{\mbox{H$\beta$}}

\newcommand{\han}{\mbox{H$\alpha$}}

\newcommand{\nii}{\mbox{[\ion{N}{2}] $\lambda$6584} $\,$}

\slugcomment{\it ApJ accepted}
\shortauthors{Comerford et al.}
\shorttitle{The Origin of Double-peaked Narrow Lines in Active Galactic Nuclei.  IV.  Association with Galaxy Mergers}

\begin{document}

\title{The Origin of Double-peaked Narrow Lines in Active Galactic Nuclei.  IV.  \\ Association with Galaxy Mergers}

\author{Julia M. Comerford\altaffilmark{1}, Rebecca Nevin\altaffilmark{1}, Aaron Stemo\altaffilmark{1}, Francisco M\"{u}ller-S\'{a}nchez\altaffilmark{1,2}, R. Scott Barrows\altaffilmark{1}, Michael C. Cooper\altaffilmark{3}, and Jeffrey A. Newman\altaffilmark{4}}

\affil{$^1$Department of Astrophysical and Planetary Sciences, University of Colorado, Boulder, CO 80309, USA}
\affil{$^2$Physics Department, University of Memphis, Memphis, TN 38152, USA}
\affil{$^3$Center for Galaxy Evolution, Department of Physics and Astronomy, University of California, Irvine, 4129 Frederick Reines Hall, Irvine, CA 92697, USA}
\affil{$^4$Pittsburgh Particle Physics, Astrophysics, and Cosmology Center, Department of Physics and Astronomy, University of Pittsburgh, Pittsburgh, PA 15260, USA}

\begin{abstract}
Double-peaked narrow emission lines in active galactic nucleus (AGN) spectra can be produced by AGN outflows, rotation, or dual AGNs, which are AGN pairs in ongoing galaxy mergers.  Consequently, double-peaked narrow AGN emission lines are useful tracers of the coevolution of galaxies and their supermassive black holes, as driven by AGN feedback and AGN fueling.  We investigate this concept further with follow-up optical longslit observations of a sample of 95 Sloan Digital Sky Survey (SDSS) galaxies that have double-peaked narrow AGN emission lines. Based on a kinematic analysis of the longslit spectra, we confirm previous work that finds that the majority of double-peaked narrow AGN emission lines are associated with outflows.  We also find that eight of the galaxies have companion galaxies with line-of-sight velocity separations $< 500$ km s$^{-1}$ and physical separations $<30$ kpc.  Since we find evidence of AGNs in both galaxies, all eight of these systems are compelling dual AGN candidates.  Galaxies with double-peaked narrow AGN emission lines occur in such galaxy mergers at least twice as often as typical active galaxies.  Finally, we conclude that at least 3\% of SDSS galaxies with double-peaked narrow AGN emission lines are found in galaxy mergers where both galaxies are resolved in SDSS imaging.
\end{abstract}  


\section{Introduction}
\label{intro}

Observational correlations between galaxies and their central supermassive black holes (e.g., \citealt{MA98.2,GE00.1,MA03.5}) indicate that supermassive black holes and galaxies are coupled as they evolve.  Active galactic nuclei (AGNs) play an important role in this coevolution.  AGNs signal that supermassive black holes are growing in mass by accreting gas from the host galaxy, and \cite{SO82.1} argues that the luminous AGN phase is when supermassive black holes build up most of their mass.  Several subsequent studies have reaffirmed this argument (e.g., \citealt{YU02.1,MA04.4,HO07.4}).

Galaxy mergers are a significant triggering mechanism for this supermassive black hole growth, because they drive inflows of gas towards the nucleus due to gravitational torques (e.g., \citealt{SA88.2}).  Ongoing galaxy mergers also bring together pairs of supermassive black holes. Inflows of gas onto these black holes create dual AGNs if both black holes are active and offset AGNs if one black hole is active \citep{KO12.1,CO14.1,BA16.1}.  Galaxy mergers also trigger bursts of star formation (e.g., \citealt{BA91.1}).

Then, star formation driven and AGN driven outflows disperse the remaining gas in the merged galaxy systems, preventing the galaxies from becoming too massive (e.g., \citealt{KI03.1,DI05.1,CR06.1,FA12.2}).  This feedback effect regulates both star formation and the growth of supermassive black holes, and helps merger remnants evolve to red, quiescent galaxies.    

Both supermassive black hole mass growth in galaxy mergers and AGN outflows can be traced by galaxy spectra with AGN emission lines that are double-peaked. In recent years, double-peaked narrow emission lines in AGN host galaxies have been studied as a population.  Several hundred double-peaked narrow emission lines have been identified in spectroscopic surveys \citep{GE07.1,CO09.1,WA09.1,XU09.1,LI10.1,SM10.1,GE12.1,BA13.1,CO13.1}, with the largest samples found in the Sloan Digital Sky Survey (SDSS).  These double-peaked line profiles can be produced by a variety of physical processes, including dual AGNs, AGN outflows, and disk rotation.  Consequently, follow-up observations are required to determine the nature of double-peaked narrow emission lines.

\begin{deluxetable*}{llllllll}
\tabletypesize{\scriptsize}
\tablewidth{0pt}
\tablecolumns{8}
\tablecaption{Summary of Observations} 
\tablehead{
\colhead{SDSS Designation} &
\colhead{Telescope/Instrument} & 
\colhead{Spectral} &
\colhead{BPT Emission} &
\colhead{Observation} &
\colhead{PA$_{\mathrm{obs,1}}$ ($^\circ$)} &
\colhead{PA$_{\mathrm{obs,2}}$ ($^\circ$)} &
\colhead{Exp.} \\
 & & Resolution & Lines Covered & Date (UT) & & & Time (s)
}
\startdata 
SDSS J000656.85+154847.9 & MMT/Blue Channel & 3800 & H$\beta$, [O III] & 2010 Nov 6 & 44.4 & 134.4 & 1080 \\
SDSS J010750.48$-$005352.9 & Gemini/GMOS-S(1x1) & 2100 & H$\beta$, [O III] & 2010 Nov 22 /  & 52.0 & 142.0 & 3600 \\
& & & & 2010 Dec 18 & & & \\
SDSS J011659.59$-$102539.1 & MMT/Blue Channel & 3800 & H$\beta$, [O III] & 2010 Nov 5 & 28.8 & 118.8 & 1080 \\
SDSS J011802.94$-$082647.2 & MMT/Blue Channel & 3800 & H$\beta$, [O III] & 2010 Nov 5 & 42.3 & 132.3 & 1260
\enddata
\tablecomments{We observed each galaxy at two position angles, PA$_{\mathrm{obs,1}}$ and PA$_{\mathrm{obs,2}}$ (given in degrees East of North), and the exposure time given is for each position angle.  \\ (This table is available in its entirety in a machine-readable form in the online journal.  A portion is shown here for guidance regarding its form and content.)}
\label{tbl-1}
\end{deluxetable*}

These follow-up observations have included high-resolution imaging to determine if multiple stellar bulges are present (e.g., \citealt{RO11.1,SH16.1}), as expected in the galaxy merger scenario for dual AGNs.  Spatially resolved spectroscopy (e.g., \citealt{CO09.3,CO12.1,GR12.2,NE16.1}) is complementary in that it can be used to constrain the kinematics of the narrow-line region (NLR) gas associated with the double peaks.  In fact, many studies have combined imaging and spatially resolved spectroscopy to build a clearer picture of the sources of the double-peaked narrow emission lines (e.g., \citealt{LI10.2,MC11.1,SH11.1,BA12.1,FU12.1,MC15.1}).

To spatially resolve and confirm dual AGNs in any of the double-peaked emission line systems, observations are required in either the radio (e.g., \citealt{RO10.1,FU11.3,TI11.1,MU15.1,SH16.1,LI18.1}) or the X-ray (e.g., \citealt{CO11.2,LI13.1,CO15.1}).  As a result of all of these multiwavelength follow-up campaigns, a picture is emerging where the majority of double-peaked emission lines are produced by the kinematics of AGN outflows.  A uniform study of a large sample of objects is needed to solidify these conclusions.

In this paper we present and analyze follow-up observations for 95 galaxies in the SDSS that have double-peaked narrow AGN emission lines in their spectra, which makes this the largest sample of such galaxies yet published in a follow-up observational paper.  We have obtained follow-up optical longlist spectroscopy for each galaxy, which we use to kinematically classify the source of the double-peaked emission line profile in each galaxy.  Using our optical longlist spectroscopy and existing imaging, we also identify eight galaxies that are in mergers and we analyze the role that they play in the larger framework of merger-driven galaxy evolution.

This paper is organized as follows.  In Section~\ref{sample}, we present the sample of 95 galaxies with double-peaked AGN emission lines in their SDSS spectra.  Section~\ref{observations} discusses our analysis of the optical longslit spectra and optical and near-infrared imaging, and our identification of galaxy pairs, including eight dual AGN candidates.  In Section~\ref{results}, we describe our results, including kinematic classifications of each galaxy and the nature of the link between galaxy mergers and double-peaked emission lines.  Finally, Section~\ref{conclusions} summarizes our conclusions.

We assume a Hubble constant $H_0 =70$ km s$^{-1}$ Mpc$^{-1}$, $\Omega_m=0.3$, and $\Omega_\Lambda=0.7$ throughout, and all distances are given in physical (not comoving) units.

\section{The Sample}
\label{sample}

\begin{figure*}[!t]
\centering
\includegraphics[height=6cm]{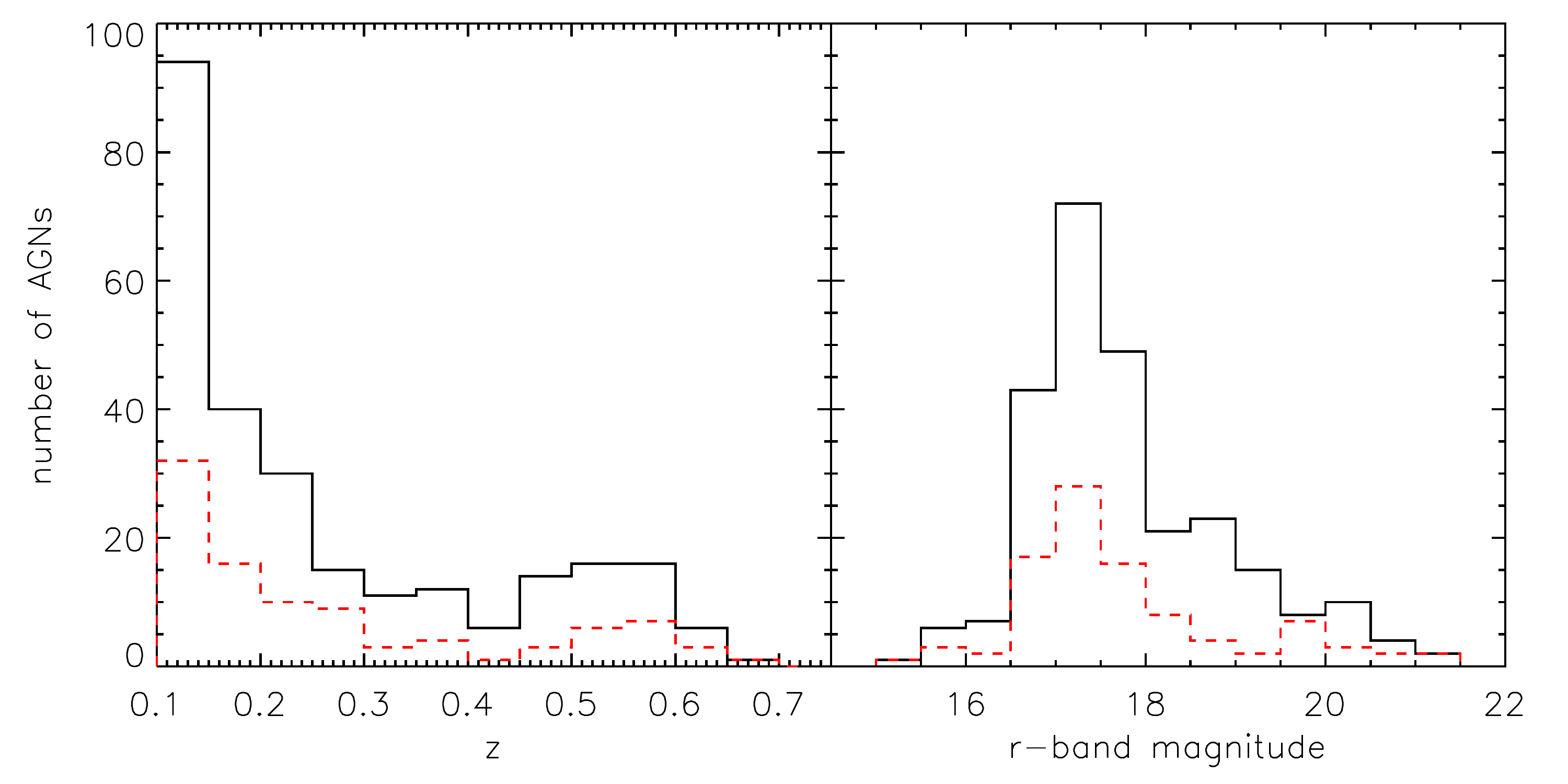}
\caption{Distributions of redshifts (left) and $r$-band magnitudes (right) for the full sample of 261 double-peaked AGNs in SDSS at $z>0.1$ (black solid histogram) and the observed sample of 95 double-peaked AGNs (red dashed histogram).  Both populations are consistent with being drawn from the same redshift distribution and $r$-band magnitude distribution.}
\label{fig:zrmag}
\end{figure*}

Our parent sample consists of SDSS AGNs with double-peaked \oiiiw emission lines in \cite{WA09.1}, \cite{LI10.1}, and \cite{SM10.1}.  Our previous work addressed the $z<0.1$ galaxies \citep{NE16.1}, and here we focus on the $z>0.1$ galaxies.  There are 261 double-peaked AGNs at $z>0.1$, and we obtained optical longslit observations for 95 of them.  The observed galaxies have redshifts $0.10 < z < 0.69$ and $r$-band magnitudes $15.4 < r < 21.2$.  The systems we observed are representative of the full $z>0.1$ catalog (Figure~\ref{fig:zrmag}), as the Kolmogorov-Smirnov probabilities are 96$\%$ (94$\%$) that the redshifts ($r$-band magnitudes) of the 95 observed AGNs and the parent sample of 261 AGNs were derived from the same distribution.

\section{Observations and Analysis}
\label{observations}

\subsection{Observations}
\label{observations}

Of the 95 $z>0.1$ galaxies in our sample, we previously observed 54 of them with the Kast Spectrograph at Lick Observatory (3 m telescope, pixel size $0\farcs78$; \citealt{MI93.3}), the Double Spectrograph (DBSP) at Palomar Observatory (5 m telescope, pixel size $0\farcs47$ for the red detector and $0\farcs39$ for the blue detector; \citealt{OK82.1}), and the Blue Channel Spectrograph at MMT Observatory (6.5 m telescope, pixel size $0\farcs29$; \citealt{AN79.1}).  We used 1200 lines mm$^{-1}$ gratings, and slit widths of $1\farcs5$, $1\farcs5$, and $1^{\prime\prime}$ for the Lick, Palomar, and MMT observations, respectively.

In addition, we obtained optical longslit observations of 41 more $z>0.1$ double-peaked AGNs with GMOS-North on the Gemini Observatory North 8 m telescope (pixel size $0\farcs073$ for $1\times1$ pixel binning and $0\farcs145$ for $1\times2$ pixel binning; \citealt{AL02.2,HO04.2}), GMOS-South on the Gemini Observatory South 8 m telescope (pixel size $0\farcs073$ for $1\times1$ pixel binning and $0\farcs146$ for $1\times2$ pixel binning), LRIS on the Keck I 10 m telescope (pixel size $0\farcs135$; \citealt{OK95.1}), and DEIMOS on the Keck II 10 m telescope (pixel size $0\farcs119$; \citealt{FA03.1}).

We observed 20 objects with Gemini/GMOS-N and 11 objects with Gemini/GMOS-S, and with both telescopes we used a 600 lines mm$^{-1}$ grating and $1^{\prime\prime}$ slit width. We observed 9 objects with Keck/LRIS; 4 of them were observed with the red and blue cameras simultaneously, and 5 were observed with the blue camera only because the red camera was broken.  We used a 1200 lines mm$^{-1}$ grating for the red camera and a 600 lines mm$^{-1}$ grating for the blue camera, and a $1^{\prime\prime}$  slit width.  We observed one object with Keck/DEIMOS, using a 1200 lines mm$^{-1}$ grating and $1^{\prime\prime}$ slit width.  

Table~\ref{tbl-1} summarizes the observations.  We observed each target with two different slit position angles.  Typically, we observed with the slit oriented along the isophotal position angle of the major axis of the object in SDSS $r$-band photometry and again with the slit at the corresponding orthogonal position angle.  In some cases we deviated from this approach due to guide star requirements or mechanical constraints on the rotation of the telescope.  

The data were reduced following standard procedures in IRAF and IDL (see \citealt{TO93.1,CO12.1,CO12.3}).

\subsection{Analysis of \oiiiw Kinematics}
\label{kinematics}

\oiiiw is a strong emission line that is an excellent tracer of the extent and kinematics of ionized gas in the NLR, and has been used in many studies of AGN photoionization and outflows (e.g., \citealt{SC03.3,GR14.1,SU17.1}).  Here, we measure kinematic parameters of the observed  \oiiiw emission in each galaxy following the approach presented in \cite{NE16.1}, who analyzed a sample of 71 galaxies with double-peaked AGN emission lines at $z<0.1$ that were selected from the same catalogs of double-peaked AGNs in SDSS that we use here (Section~\ref{sample}).  The \cite{NE16.1} sample does not overlap with the sample presented here. The full details of the kinematic analyses are given in \cite{NE16.1}, and we summarize the approach below.

First, we fit a Gaussian to the stellar continuum along the spatial position axis of each longslit spectrum, and measure the full width at half maximum of the spatial extent of the continuum (FWHM$_\mathrm{cont}$).  For each spatial row of the longslit spectrum that is within FWHM$_\mathrm{cont}$, we determine how many Gaussians are required to best fit the \oiiiw emission.  To determine the number of Gaussians required for the fit, we use the Akaike Information Criterion (AIC), which is a least squares statistic that penalizes extra free parameters \citep{AK74.1}.  The fraction of the spatial rows of the longslit spectrum within FWHM$_\mathrm{cont}$ that are best fit by more than two Gaussians is used in our kinematic classification of the galaxies (Section~\ref{classify}).

Then, we measure the spatial extent of the \oiiiw emission.  For each spatial row of the longslit spectrum, we use the AIC to determine whether the \oiiiw spectrum is better fit by a line (representing the background; 2 free parameters) or by a Gaussian and a line (representing an emission line and the background; 5 free parameters in total).  We define the spatial extent of the \oiiiw emission as the spatial extent along the slit for which each spatial row's spectrum is best fit by a Gaussian and a line.  For each row of the longslit spectrum that is within this spatial extent, we fit the \oiiiw emission line first with one Gaussian, and then with two Gaussians.  For the single-Gaussian fit, we measure the line-of-sight velocity difference between the central velocity of the Gaussian fit and the systemic velocity of the galaxy in each row, and we report the highest measured line-of-sight velocity difference $V_r$.  We also measure the dispersion of each component of the double Gaussian fit in each row, and we report the highest measured dispersions $\sigma_1$ and $\sigma_2$. 

Further, we determine the position angle of the maximum extent of \oiiiw on the sky, by measuring the spatial positions of \oiiiw along each slit position angle as in \cite{CO12.1}.  We compare this to PA$_{gal}$, the isophotal position angle of the major axis of the galaxy from SDSS $r$-band photometry.  If PA$_\mathrm{[OIII]} =$ PA$_\mathrm{{gal}}$ to within $20^\circ$ error \citep{NE16.1}, then we classify the \oiiiw emission as aligned with the plane of the galaxy. 

Finally, we measure the asymmetry $A$ of the \oiiiw emission line profile in each longslit spectrum, following the approach of \cite{WH85.1} and \cite{LI13.2}, and where positive asymmetry values indicate redshifted emission-line wings and negative asymmetry values indicate blueshifted emission-line wings.  We consider the asymmetries measured from the two position angle observations, and we report the $A$ that has the larger absolute value.  Then, we define an emission line profile as symmetric if the absolute value of its asymmetry falls below the 95\% confidence interval around the mean asymmetry of the sample.  Consequently, symmetric profiles are defined by $|A|<0.19$, as in \cite{NE16.1}.

The kinematic parameters are given in Table~\ref{tbl-kinematics}, and these are the parameters that we will use to kinematically classify the source of the double-peaked emission lines in each galaxy.  As described in Section~\ref{classify}, the main classifications are rotation dominated, outflow, and ambiguous.

\begin{deluxetable*}{lccllllccc}
\tabletypesize{\scriptsize}
\tablewidth{0pt}
\tablecolumns{10}
\tablecaption{Kinematic Classifications of the 95 Galaxies} 
\tablehead{
\colhead{SDSS Name} & 
\colhead{PA$_{\mathrm{obs}}$} & 
\colhead{Num. Rows Fit} & 
\colhead{Single Gauss} & 
\colhead{Double Gauss} & 
\colhead{Double Gauss} & 
\colhead{PA$_\mathrm{{gal}}$} & 
\colhead{PA$_\mathrm{[OIII]}$}  &
\colhead{$A$} &
\colhead{Kinematic} \\
& & by $>2$ Gaussians & $V_r$ [km $\mathrm{s}^{-1}$] &  
$\sigma_1$ [km $\mathrm{s}^{-1}$] & $\sigma_2$ [km $\mathrm{s}^{-1}$] & & & & Classification 
}
\startdata
J0006+1548 & 44.4 & 4/11 & $-231.5^{+27.1}_{-7.2}$ & $232.7^{+37.6}_{-18.8}$ & $173.0^{+59.5}_{-26.2}$ & 44.4 & $172.6 \pm 3.9$ & 0.08 & Ambiguous \\
& 134.4 & 4/11 & $-241.9^{+25.1}_{-9.5}$ & $168.4^{+277.5}_{-0.0}$ & $172.7^{+77.2}_{-29.1}$ &  &  & 0.11 & \\
J0107$-$0053 & 52.0 & 22/43 & $305.6^{+23.1}_{-11.1}$ & $271.2^{+39.0}_{-43.0}$ & $222.4^{+13.9}_{-35.3}$ & 52.0 & $176.6 \pm 3.8$ & -0.35 & Ambiguous \\
& 142.0 & 16/41 & $380.9^{+47.9}_{-71.0}$ & $353.2^{+224.6}_{-92.7}$ & $367.6^{+127.6}_{-105.7}$ & & & -0.55 & \\
J0116$-$1025 & 28.8 & 10/13 & $33.7^{+6.0}_{-0.6}$ & $275.0^{+78.2}_{-0.6}$ & $106.6^{+78.2}_{-55.7}$ & 118.8 & $117.0 \pm 2.6$ & 0.19 & Rotation Dominated \\
& 118.8 & 4/13 & $-185.2^{+71.4}_{-0.5}$ & $260.8^{+69.4}_{-0.2}$ & $258.5^{+14.8}_{-22.3}$ & & & -0.27 & + Disturbance \\
J0118$-$0826 & 42.3 & 9/15 & $-310.5^{+40.2}_{-8.0}$ & $218.2^{+108.4}_{-8.0}$ & $338.2^{+210.1}_{-65.3}$ & 42.3 & $25.7 \pm 3.4$ & 0.18 & Rotation Dominated \\
& 132.3 & 4/13 & $-296.4^{+448.7}_{-15.2}$ & $220.3^{+208.1}_{-15.8}$ & $331.3^{+49.0}_{-67.5}$ &  &  & 0.06 & + Obscuration 
\enddata
\label{tbl-kinematics}
\tablecomments{Column 1: Galaxy name. Column 2: Observed position angle, in degrees East of North. Column 3: Number of spatial rows of emission in the longslit spectrum that are best fit by $>2$ Gaussians within the FWHM of the continuum.  Column 4: Line-of-sight velocity difference between the velocity derived from a single-Gaussian fit to the emission line profile and the systemic velocity of the galaxy. Columns 5 and 6: Dispersion of each component of a double-Gaussian fit to the emission line profile. Column 7: Position angle of the major axis of the galaxy in $r$-band, in degrees East of North. Column 8: Position angle of the maximum extent of \oiiiw on the sky, in degrees East of North. Column 9: Asymmetry of the emission line profile, where positive asymmetry values indicate redshifted emission-line wings and negative asymmetry values indicate blueshifted emission-line wings. Column 10: Kinematic classification of the galaxy. \\ (This table is available in its entirety in a machine-readable form in the online journal.  A portion is shown here for guidance regarding its form and content.)}
\end{deluxetable*}

\subsection{Analysis of Galaxy Pairs}
\label{pairs}

We searched the 95 galaxies for those with companion galaxies that have line-of-sight velocity separations $| \Delta v | < 500$ km s$^{-1}$ and projected physical separations $\Delta x <30$ kpc, where Section~\ref{longslit} and Section~\ref{imaging} describe how $\Delta v$ and $\Delta x$ are measured, respectively.  We selected the  $| \Delta v | < 500$ km s$^{-1}$ and $\Delta x <30$ kpc criteria to match those of \cite{EL08.1}, for ease of later comparison.

We found that eight galaxies with double-peaked AGN emission lines have companion galaxies within $| \Delta v | < 500$ km s$^{-1}$ and $\Delta x <30$ kpc (Figure~\ref{fig:dualoffset}).  Five of the galaxies have SDSS spectra that are classified as Type 1 AGNs (SDSS J0952+2552, SDSS J1157+0816, SDSS J1248$-$0257, SDSS J1541+2036, SDSS J1610+1308) and three have SDSS spectra that are classified as Type 2 AGNs (SDSS J1245+3723, SDSS J1301$-$0058, SDSS J1323+0308).  

As a further test of whether two galaxies in a pair are indeed related, we can neglect the $| \Delta v |$ cutoff and estimate the probability of the chance projection of two unassociated galaxies on the sky.  We carry out this calculation for the galaxy pairs (SDSS J1157+0816, SDSS J1245+3723, SDSS J1323+0308) where both galaxies are detected in the SDSS photometric catalog, which has a limiting $r$-band magnitude of 22.0 \citep{AL15.2}.  The larger separation pairs have greater probabilities of being chance projections of unrelated galaxies, and this subsample does include the largest separation pair (SDSS J1323+0308, with a separation of $28.5$ kpc).  The faintest galaxy in each pair has an $r$-band magnitude of (19.66, 20.41, 19.29) for (SDSS J1157+0816, SDSS J1245+3723, SDSS J1323+0308), respectively.  For each galaxy pair, we calculate the surface density SDSS galaxies out to this $r$-band magnitude limit.  For the example of the faintest galaxy in our pair sample, the $r=20.41$ galaxy SDSS J1245+3723, the surface density out to $r=20.41$ is 209 galaxies deg$^{-2}$.  We calculate the probability of chance alignment for each galaxy pair using the surface density, redshift, and the solid angle subtended by the galaxy pair separation that we measured (Table~\ref{tbl-comp}).  We find that the probability of chance alignment is ($2 \times 10^{-4}$, $1 \times 10^{-4}$, $9 \times 10^{-4}$) for the galaxy pair systems (SDSS J1157+0816, SDSS J1245+3723, SDSS J1323+0308), respectively. Consequently, the probability is 0.001 that there are one or more chance projections in this subsample.  The likelier scenario is that the galaxies in these pairs are in fact associated with one another. 

The subsections that follow focus on the longslit spectra and the images of these eight merging galaxy systems.  We note that many other galaxies in our sample of 95 galaxies have companions within 30 kpc that were resolved with near-infrared imaging \citep{RO11.1,FU12.1,MC15.1}, but since the companions do not have spectroscopic redshifts they cannot yet be confirmed as associated with the primary galaxies.  

\subsubsection{Longslit Spectral Analysis}
\label{longslit}

\begin{figure*}
\centering
\subfigure{\includegraphics[width=7.in]{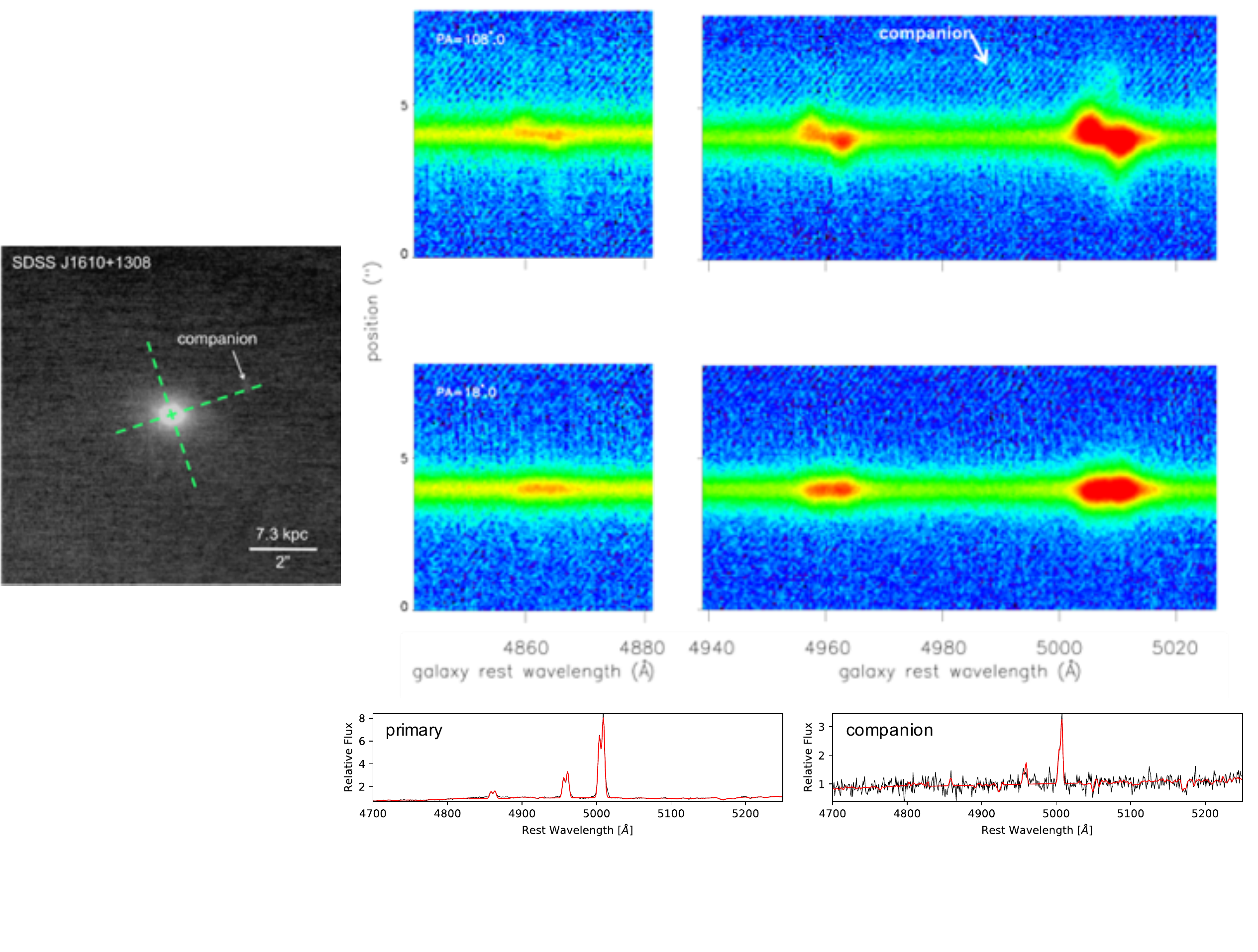}}
\caption{Imaging and spectra of the eight galaxies with double-peaked AGN emission lines that have companion galaxies within $| \Delta v | < 500$ km s$^{-1}$ and $\Delta x <30$ kpc. Left: Images of the galaxies, from {\it HST}, SDSS, or NIRC2. The dashed green lines show the orientations of the two longslit positions.  In all images, north is up and east is left.  Right: Segments of the two-dimensional long-slit spectra, and the one-dimensional spectra extracted for each galaxy and its companion.  The one-dimensional spectra are normalized by dividing each spectrum by the mean of the continuum flux, and the red curves show the fits to the data.  The telluric A-band absorption is masked during the spectral fits, as shown by the grey bands.}
\label{fig:dualoffset}
\end{figure*}

Here we determine the redshifts and emission line fluxes for the galaxies and their companions.  We note that for SDSS J1301$-$0058, the slit is slightly misaligned with the companion because the longslit observations were taken before the image revealing the companion was published (Figure~\ref{fig:dualoffset}).  Since the $1^{\prime\prime}$ slit covered the center of the companion, the derived redshifts and emission line fluxes for the companion are still appropriate to use. 

Using the longslit spectra, we extracted the redshifts of the galaxies with double-peaked AGN emission lines and the redshifts of the companion galaxies.  We used a penalized pixel-fitting software (pPXF; \citealt{CA04.2}) to simultaneously fit emission-line and absorption-line templates to the observed spectra, where we masked the telluric A-band absorption.  This yielded the absorption-line redshifts $z_{DPAGN}$ and $z_{companion}$, which enabled us to measure the line-of-sight velocity difference $\Delta v = v_{DPAGN}-v_{companion}$.  Table~\ref{tbl-comp} shows the redshifts and the line-of-sight velocity differences. 

We also used Baldwin-Phillips-Terlevich (BPT) emission line diagnostics \citep{BA81.1,KE06.1} to identify the source of any ionized emission present in the companion galaxy.  To do so, we fit Gaussians to the \hbn, \oiiiwn, \han, and \nii emission lines to determine their fluxes.  The resultant BPT classifications are given in Table~\ref{tbl-comp}.  

\begin{deluxetable*}{llllllll}
\tablewidth{0pt}
\tablecolumns{8}
\tablecaption{Galaxies with Double-peaked AGN Emission Lines and Companion Galaxies within $| \Delta v | < 500$ km s$^{-1}$ and $\Delta x <30$ kpc} 
\tablehead{
\colhead{SDSS Name} & 
\colhead{$z_{DPAGN}$} &
\colhead{$z_{companion}$} &
\colhead{$v_{DPAGN}-$ } &
\colhead{Distance to} & 
\colhead{Distance to} &
\colhead{Companion} &
\colhead{System} \\
 & & & $v_{companion}$ & Companion & Companion & BPT & Classification \\
 & & & (km s$^{-1}$) & ($^{\prime\prime}$) & (kpc) & Classification &  
 }
\startdata 
J0952+2552  & $0.33887 \pm 0.00030$ & $0.33888 \pm 0.00199$ & $\phantom{0}-3 \pm 130$ & $1.00 \pm 0.01^a$ & $\phantom{0}4.82 \pm 0.05$ & Seyfert & Candidate Dual AGNs \\ 
J1157+0816 & $0.20133 \pm 0.00042$ & $0.20201 \pm 0.00385$ & $-170 \pm 262$ & $2.52 \pm 0.01$ & $\phantom{0}8.35 \pm 0.01$ & Seyfert & Candidate Dual AGNs \\ 
J1245+3723 & $0.27910 \pm 0.00062$ & $0.27836 \pm 0.00125$ & $\phantom{-}173 \pm 162$ & $1.38 \pm 0.01$ & $\phantom{0}5.83 \pm 0.02$ & Composite & Candidate Dual AGNs \\ 
J1248$-$0257 & $0.48675 \pm 0.00161$ & $0.48774 \pm 0.00145$ & $-199 \pm 333$  & $0.53^b$ & $\phantom{0}3.2$ & Seyfert & Candidate Dual AGNs \\ 
J1301$-$0058 & $0.24544 \pm 0.00030$ & $0.24514 \pm 0.00137$ & $\phantom{-}\phantom{0}72 \pm 109$ & $1.40^c$ & $\phantom{0}5.4$ & Seyfert & Candidate Dual AGNs \\ 
J1323+0308 & $0.26953 \pm 0.00114$ & $0.26978 \pm 0.00121$ & \phantom{0}$-59 \pm 278$ & $6.90 \pm 0.01$ & $28.46 \pm 0.02$ &  Seyfert & Candidate Dual AGNs \\ 
J1541+2036 & $0.50795 \pm 0.00137$ & $0.50851 \pm 0.00131$ & $-112 \pm 281$ & $2.00^b$ & $12.2$ & Seyfert & Candidate Dual AGNs \\ 
J1610+1308 & $0.22868 \pm 0.00021$ & $0.22925 \pm 0.00053$ & $-138 \pm 61$ & $2.35^b$ & $\phantom{0}8.6$ & Seyfert & Candidate Dual AGNs 
\enddata 
\tablecomments{DPAGN denotes double-peaked AGN. \\
$^a$ measured from {\it HST}/WFC3/F160W imaging \citep{CO15.1} \\
$^b$ measured from NIRC2 imaging \citep{RO11.1} \\
$^c$ measured from {\it HST}/ACS/F550M imaging \citep{FU12.1}}
\label{tbl-comp}
\end{deluxetable*}

We found two companions with Seyfert line flux ratios, and one with composite line flux ratios.  For five of the companions, the wavelength ranges of the longslit observations only covered \hb and \oiiiw (Table~\ref{tbl-1}), preventing us from completing the full BPT diagnostic.  For these galaxies, we conservatively used $\oiiiwn/\hbn >3$ to distinguish Seyferts from other line-emitting galaxies (e.g., \citealt{YA06.1,CO09.1}) and all five galaxies met this Seyfert criterion.  As we discuss in Section~\ref{dual}, none of these companions should be considered as confirmed AGNs without radio or X-ray detections.  

It is important to note that a typical NLR can be spatially extended up to $\sim10$ kpc (e.g., \citealt{SU17.1}), and that many of the galaxy pairs have separations that are less than 10 kpc (SDSS J0952+2552, SDSS J1157+0816, SDSS J1245+3723, SDSS J1248$-$0257, SDSS J1301$-$0058, SDSS J1610+1308).  Consequently, emission from the AGN in the primary galaxy could spatially overlap with the companion galaxy and produce Seyfert-like line flux ratios that are not in fact associated with an AGN located in the companion galaxy.  For this reason, we cannot confirm any galaxy pair as hosting dual AGNs with these longslit data alone.

\subsubsection{Imaging Analysis}
\label{imaging}

We also measured the spatial separation between each galaxy and its companion, using the SDSS $z$ band image (central wavelength 8932 $\mathrm{\AA}$) since it is dominated by stellar continuum.  We ran Source Extractor \citep{BE96.1} with a detection threshold of $5\sigma$ above the background to determine each galaxy's centroid, and then we used the pixel scale to convert the pixel separation into arcseconds.  Two of the galaxies were imaged with the {\it Hubble Space Telescope} ({\it HST}) and three were imaged with the NIRC2 near-infrared camera on the Keck II telescope, and for these five galaxies we used the spatial separations measured from those data \citep{RO11.1,FU12.1,CO15.1}.  Table~\ref{tbl-comp} reports the spatial separation between each galaxy and its companion.

To determine the mass ratio of the merging galaxies, we compared the luminosity of the galaxy with double-peaked emission lines to the luminosity of its companion galaxy.  We determined galaxy luminosities using the SDSS $z$ band images, except in the cases where {\it HST} or NIRC2 images are available.  For the SDSS $z$ band and {\it HST} images, we modeled the galaxies with S{\'e}rsic profiles and a uniform sky background using GALFIT V3.0 \citep{PE10.1}.  GALFIT outputs the integrated magnitudes of the galaxies.  Then, we converted the integrated magnitudes to luminosities and determined the luminosity ratio.  For the galaxies imaged with NIRC2, we took the apparent magnitudes reported in \cite{RO11.1} and converted them to luminosities.  The luminosity ratios of the merging galaxies are given in Table~\ref{tbl-prop}, where we denote the more luminous stellar bulge with ``1" and the less luminous stellar bulge with ``2."

\begin{figure*}[!t]
\centering
\vspace{-.1in}
\subfigure{\includegraphics[width=7.in]{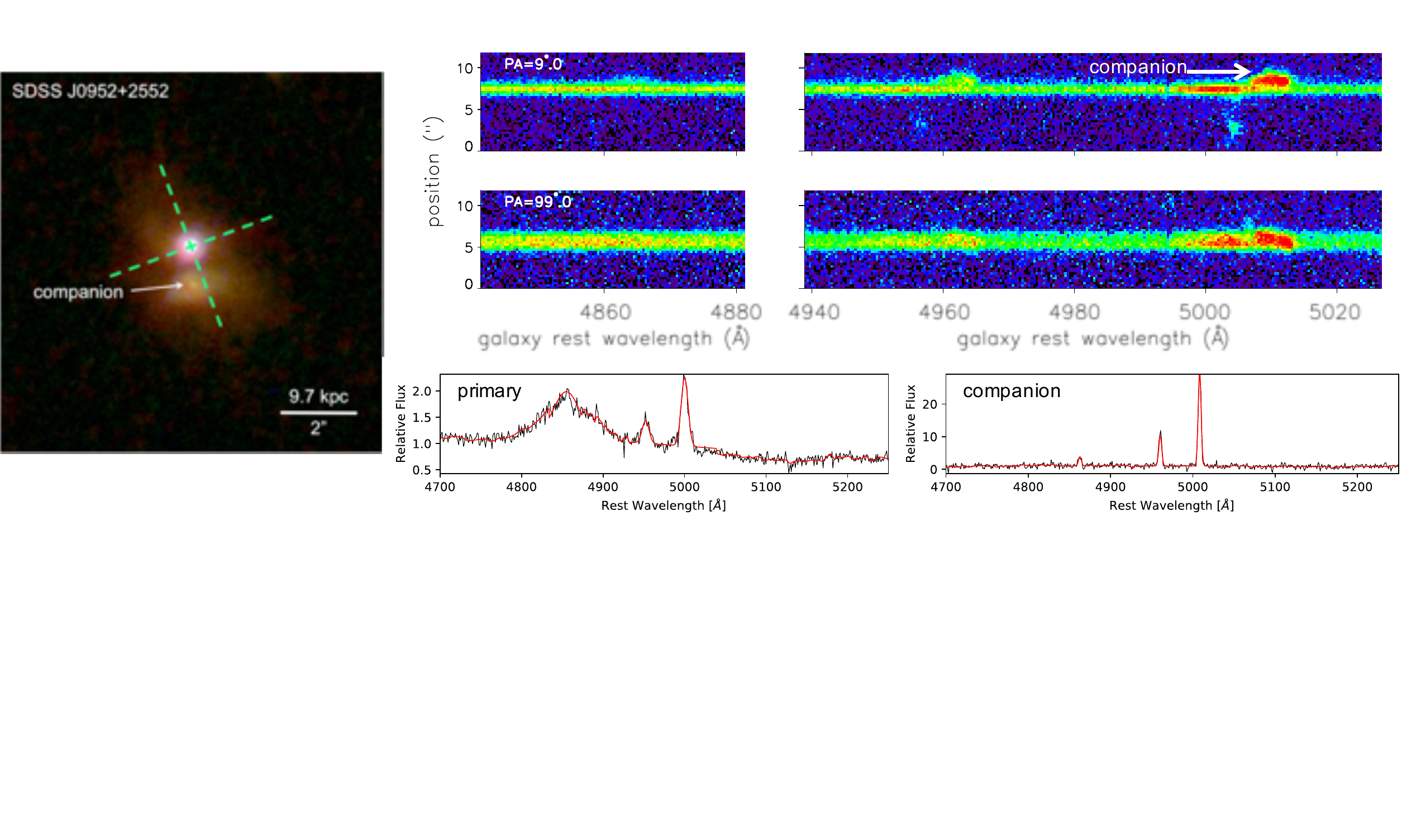}}
\subfigure{\includegraphics[width=7.in]{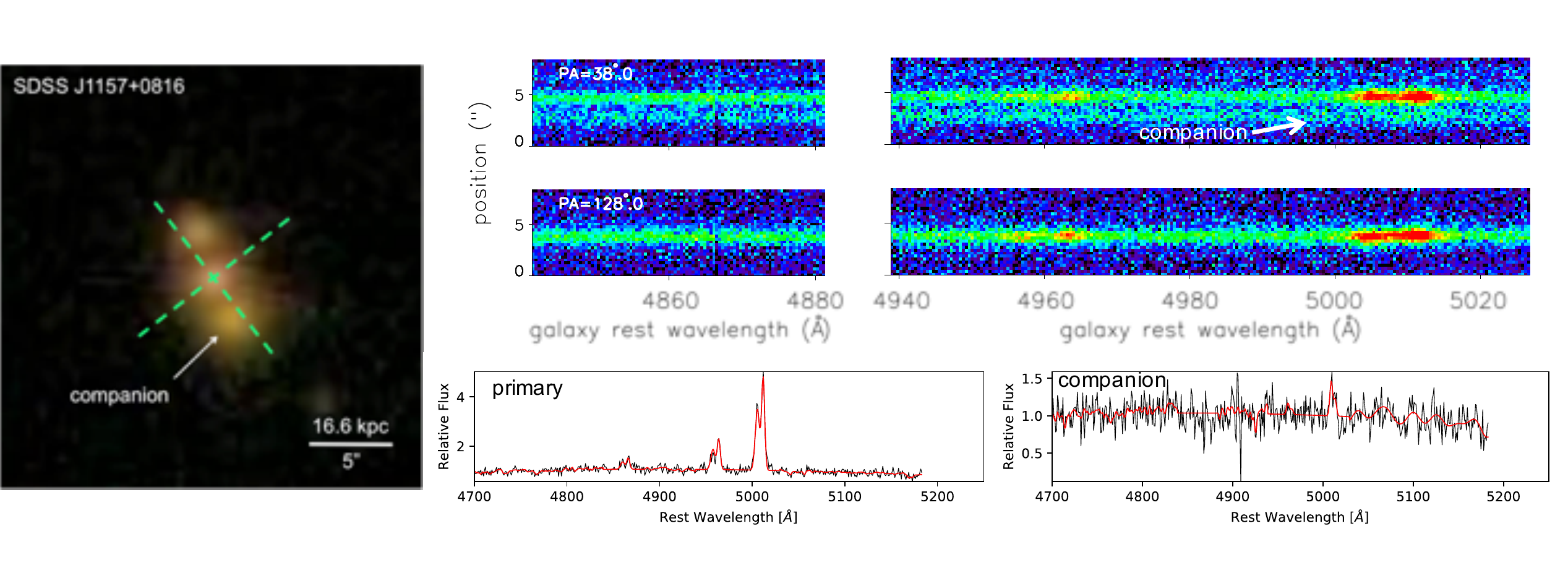}}
\subfigure{\includegraphics[width=7.in]{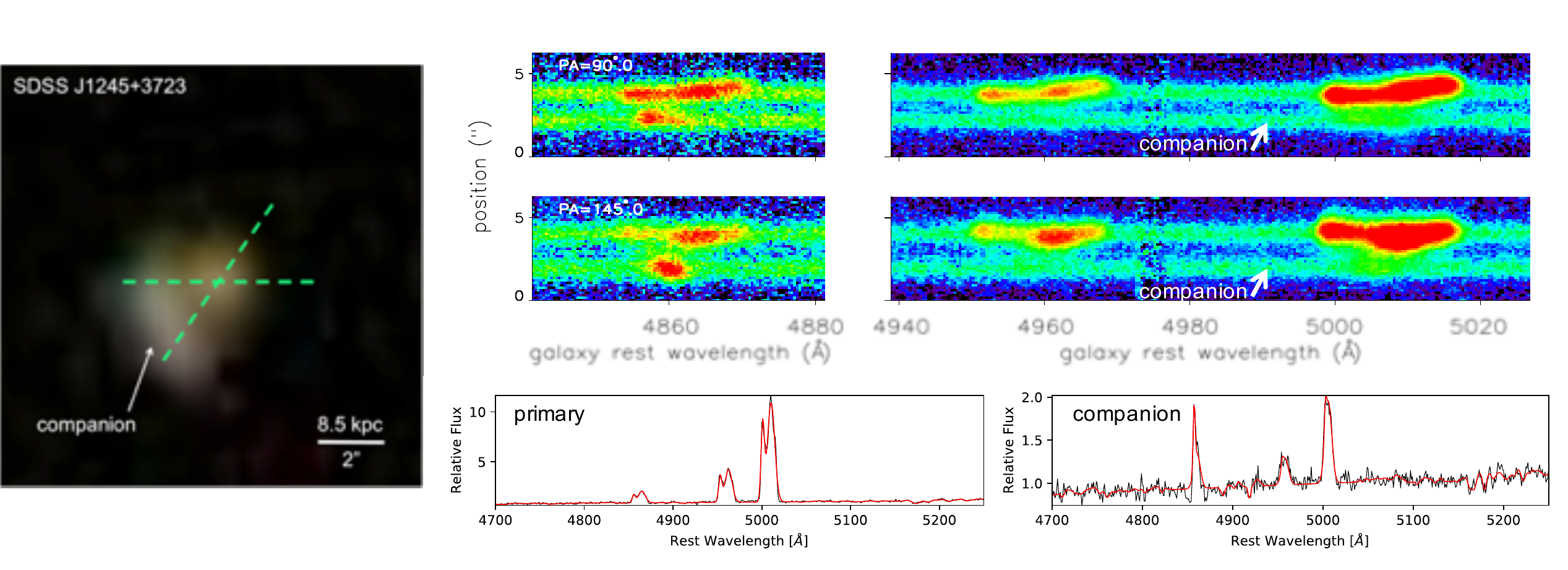}}
\vspace{.5in}
\end{figure*}

\begin{figure*}
\centering
\subfigure{\includegraphics[width=7.in]{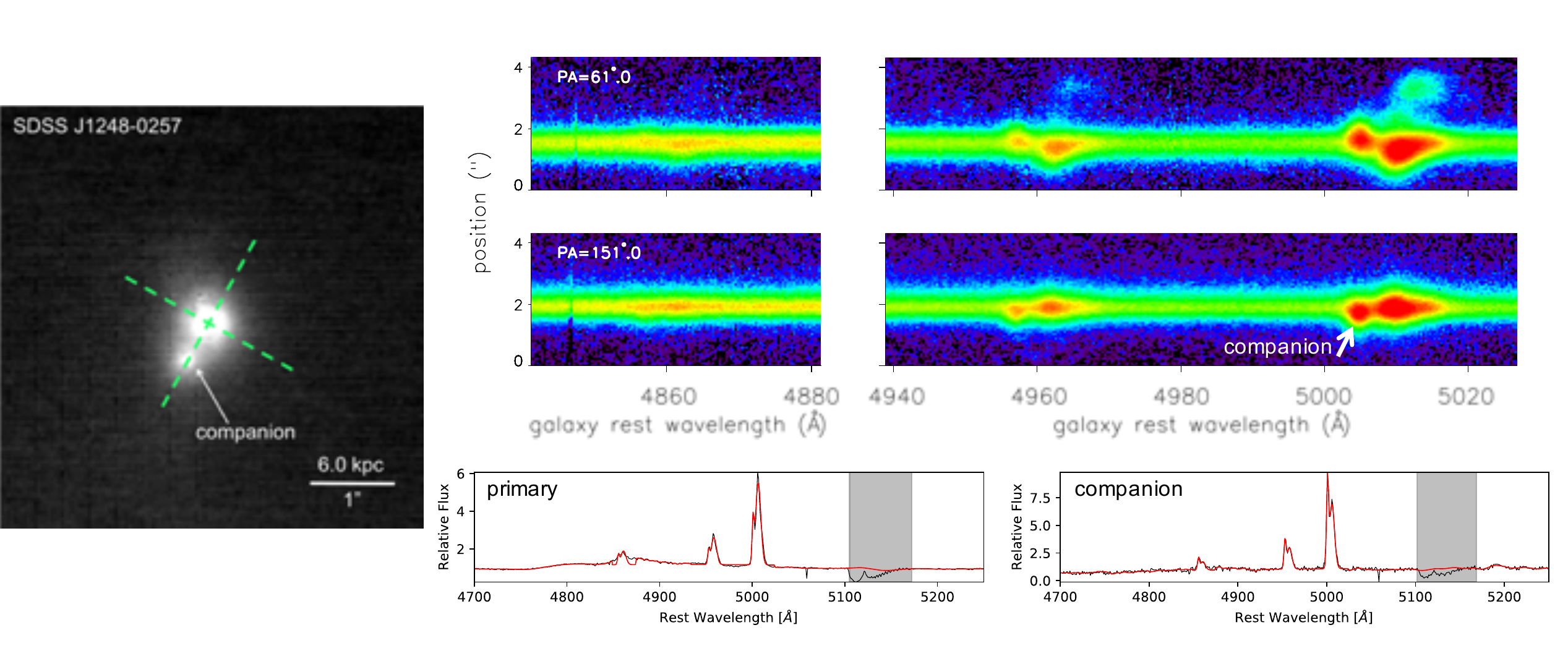}}
\subfigure{\includegraphics[width=7.in]{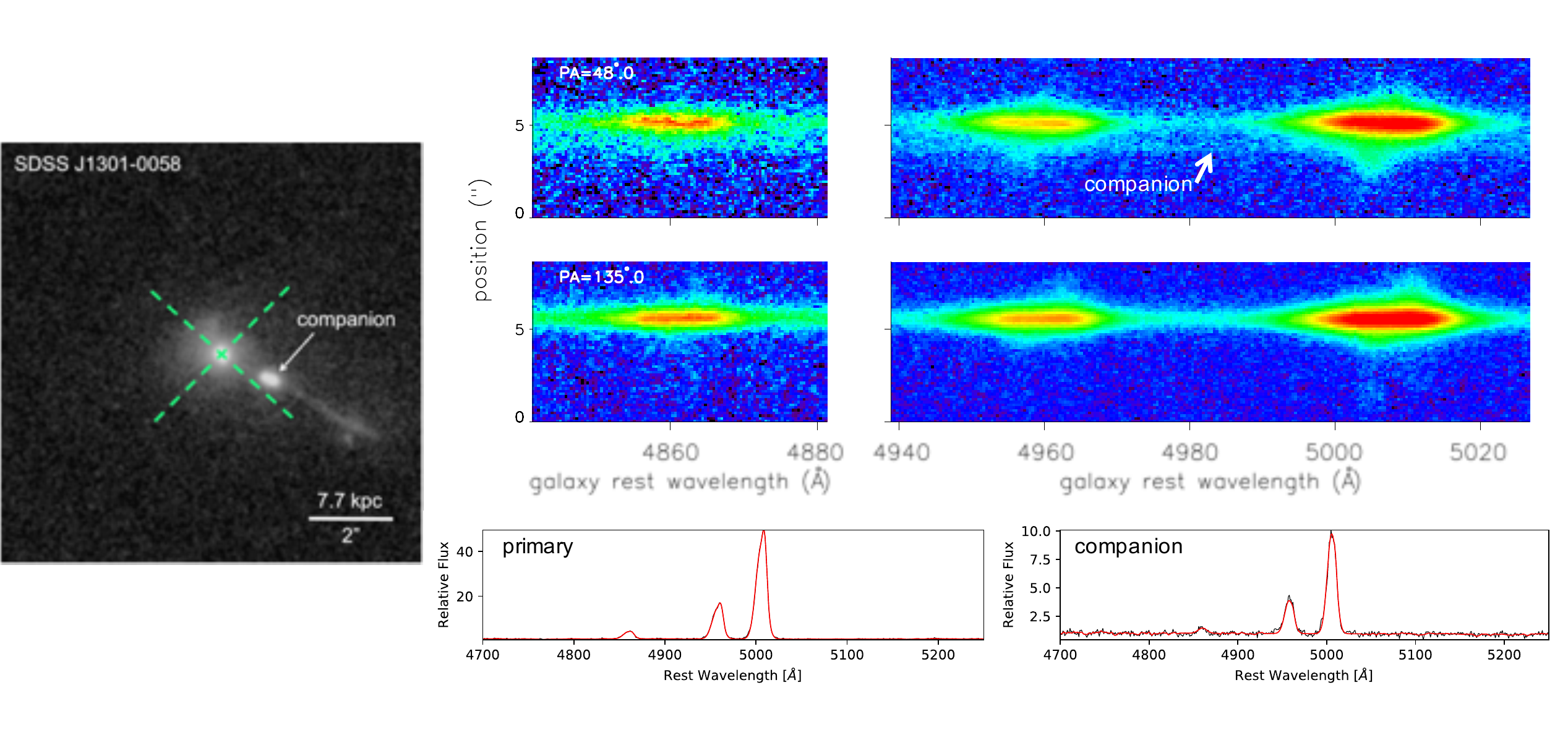}}
\vspace{2.in}
\end{figure*}

\begin{figure*}
\centering
\subfigure{\includegraphics[width=7.in]{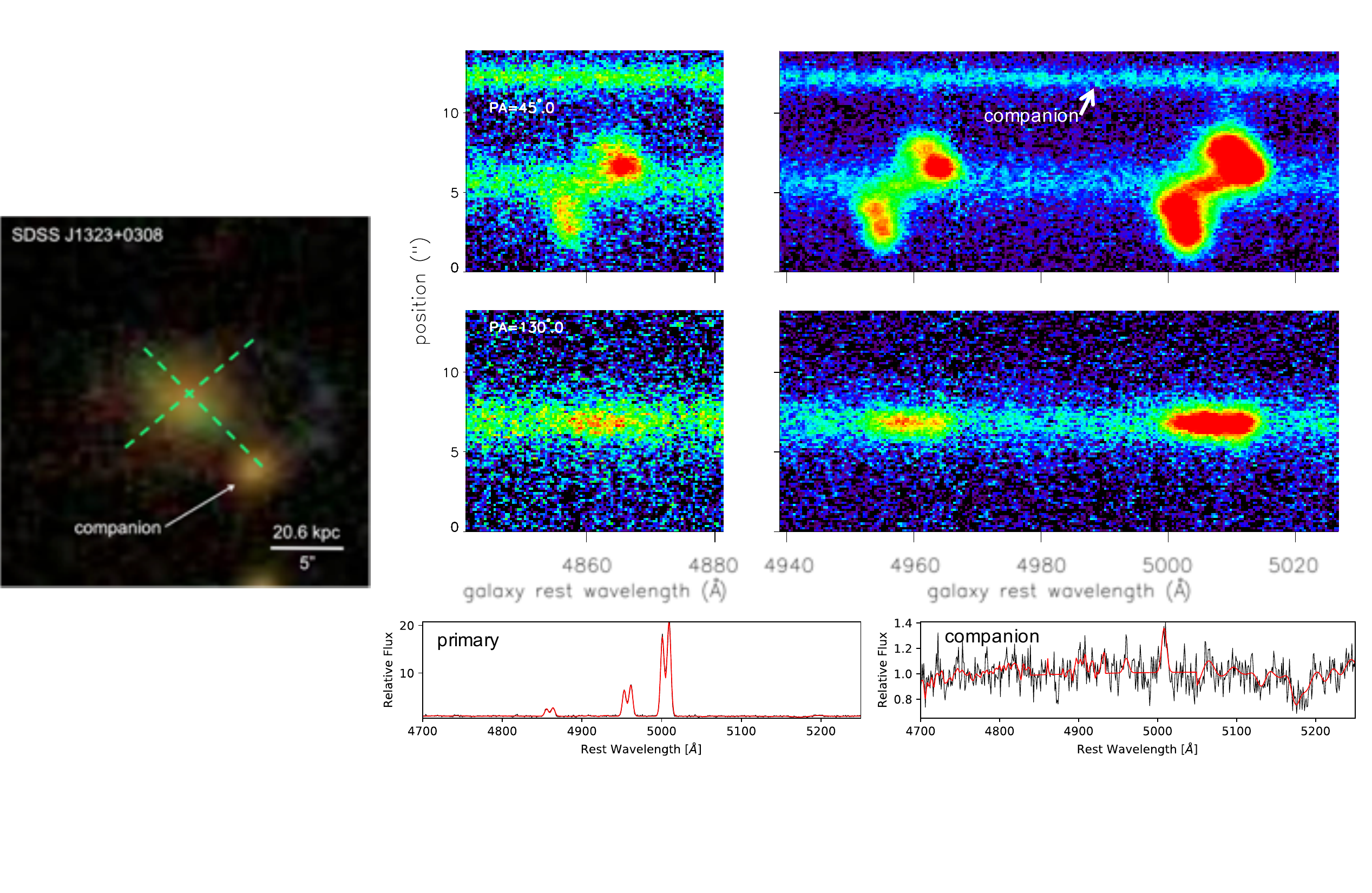}}
\subfigure{\includegraphics[width=7.in]{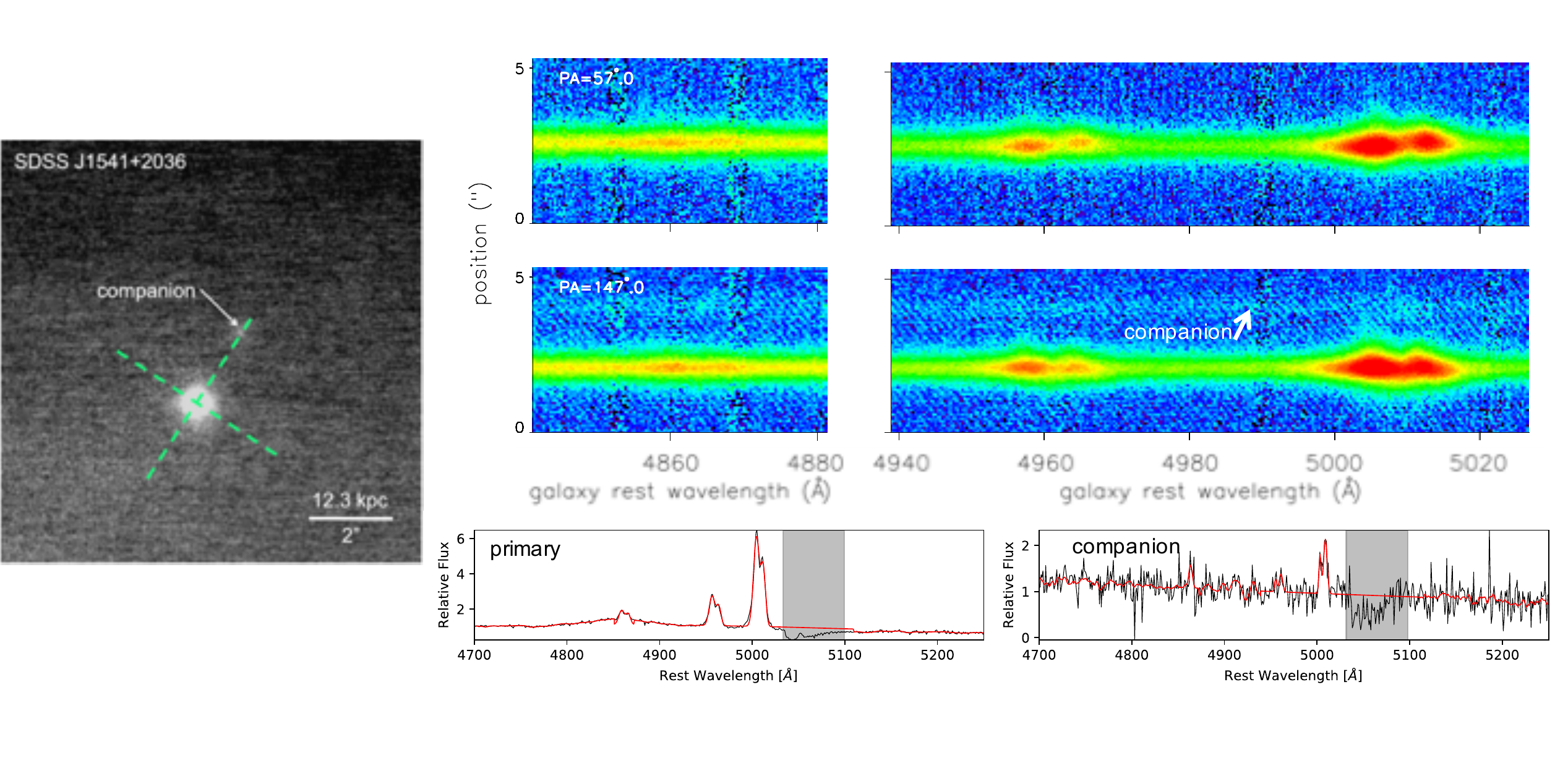}}
\end{figure*}

\begin{deluxetable*}{lllll}
\tablewidth{0pt}
\tablecolumns{5}
\tablecaption{Properties of the Merging Galaxy Systems} 
\tablehead{
\colhead{SDSS Name} & 
\colhead{Galaxy$_1$: Galaxy$_2$} &
\colhead{Merger Ratio} &
\colhead{$\log (L_\mathrm{bol,1}/$} & 
\colhead{$\log(f_\mathrm{Edd,1}/$} \\
 & & $(L_{*,1}/L_{*,2})$ & $L_\mathrm{bol,2})$ & $f_\mathrm{Edd,2})$ 
 }
\startdata  
J0952+2552  & DPAGN: companion & $2.0 \pm 0.18^a$ & $0.15$ & $-0.15$ \\
J1157+0816 & companion: DPAGN & $1.5 \pm 0.24$ & $-1.16$ & $-1.34$ \\
J1245+3723 & DPAGN: companion & $2.2 \pm 0.31$ & $1.22$ & $0.88$  \\
J1248$-$0257 & companion: DPAGN & $1.4 \pm 0.11^b$ & $-0.85$ & $-1.00$  \\ 
J1301$-$0058 & DPAGN: companion & $3.0 \pm 0.28^c$ & $0.63$ & $0.15$ \\
J1323+0308 & DPAGN: companion & $5.6 \pm 0.73$ & $2.20$ & $1.45$  \\
J1541+2036 & DPAGN: companion & $3.0 \pm 0.33^b$ & $2.12$ & $1.64$  \\
J1610+1308 & DPAGN: companion & $48.3 \pm 5.2^b$ & $1.97$ & $0.28$ 
\enddata 
\tablecomments{Galaxy$_1$ is the more luminous galaxy and Galaxy$_2$ is the less luminous galaxy. The errors on $L_\mathrm{bol,1}/L_\mathrm{bol,2}$ and $f_\mathrm{Edd,1}/f_\mathrm{Edd,2}$ are 0.54 dex. \\
$^a$ measured from {\it HST}/WFC3/F160W imaging \citep{CO15.1} \\
$^b$ measured from NIRC2 imaging \citep{RO11.1} \\
$^c$ measured from {\it HST}/ACS/F550M imaging}
\label{tbl-prop}
\end{deluxetable*}

\section{Results}
\label{results}

\subsection{Most Double Peaks Are Produced by Outflows}
\label{classify}

Here, we use the \oiiiw emission measurements from Section~\ref{kinematics} to kinematically classify each galaxy using the classification scheme of \cite{NE16.1}.  The three main classifications are as follows.

Rotation Dominated.  We classify a system as rotation dominated if it has Keplerian rotation in the plane of the galaxy, exhibited by $V_r < 400$ km s$^{-1}$, $\sigma_1<500$ km s$^{-1}$, $\sigma_2 <500$ km s$^{-1}$, and \oiiiw emission that is aligned with the plane of the galaxy (e.g., \citealt{OS06.1}).  If the emission line profile is symmetric ($|A|<0.19$), then we classify the system as ``Rotation Dominated + Obscuration" since dust obscuring a rotating disk could explain a symmetric double-peaked emission line profile (e.g., \citealt{SM12.1}).  If the emission line profile is asymmetric ($|A|>0.19$), then we classify the system as ``Rotation Dominated + Disturbance" since nuclear bars, spiral arms, dual AGNs, or other dynamical disturbances could produce an asymmetric double-peaked emission line profile (e.g., \citealt{SC97.2,DA09.1,BL13.1}).

Outflow. For a system to be classified an outflow, we apply the conservative criteria of $V_r > 400$ km s$^{-1}$, $\sigma_1 > 500$ km s$^{-1}$, or $\sigma_2 > 500$ km s$^{-1}$ (e.g., \citealt{DA06.1,MU11.1,FI13.1}).  Then, if more than half of the spatial rows of the longslit spectra within FWHM$_\mathrm{cont}$ have \oiiiw emission that is best fit by more than two Gaussians, we classify the system as ``Outflow Composite".  Otherwise, the system is ``Outflow".  The ``Outflow Composite" class can be explained by an outflow that includes many different gas clouds with their own distinct velocities.

Ambiguous.  If a galaxy fits neither the rotation dominated nor the outflow classification, then it is ambiguous.  These are the systems with $V_r < 400$ km s$^{-1}$, $\sigma_1 < 500$ km s$^{-1}$, $\sigma_2 < 500$ km s$^{-1}$, and \oiiiw emission that is not aligned with the plane of the galaxy.  Possible explanations for the emission line profiles of these galaxies include a counter-rotating disk, inflowing gas, weak outflows, and dual AGNs.

Table~\ref{tbl-kinematics} shows the results of our kinematic classifications. We find that the galaxies are classified as follows: $6\%$ ($6/95$) are Rotation Dominated + Obscuration, $9\%$ ($9/95$) are Rotation Dominated + Disturbance, $21\%$ ($20/95$) are Outflows, $33\%$ ($31/95$) are Outflow Composite, and $30\%$ ($29/95$) are Ambiguous.  

First, we consider the galaxies classified as Rotation Dominated + Obscuration or Rotation Dominated + Disturbance.  Simulations of the emission line profiles of AGNs in galaxy mergers have shown that a Rotation Dominated + Disturbance profile can be associated with dual AGNs, where a rotating disk around one AGN is disturbed by the second AGN \citep{BL13.1}.  In contrast, dual AGNs are unlikely to produce the symmetry of the emission lines in the Rotation Dominated + Obscuration classification; instead, these symmetric emission line profiles are most likely associated with obscured rotating disks (e.g., \citealt{SM12.1,BL13.1}). 

Next, we examine the galaxies classified as Outflow or Outflow Composite.  Outflow galaxies are the archetypal outflows that have a single redshifted component and a single blueshifted component, while Outflow Composite galaxies have evidence of three or more emission knots that are moving at distinct velocities.  This complex structure can be due to, e.g., outflowing gas shocking as it encounters the interstellar medium, or multiple different outflows (e.g., \citealt{CE02.1,CR05.1,CO17.2}).  We note that the number of galaxies classified as Outflow Composite is a lower limit, since higher signal-to-noise follow-up spectra may reveal additional Gaussian components that transform an Outflow classification into an Outflow Composite classification.  Therefore, we do not draw a significant distinction between Outflow and Outflow Composite classifications here.

Finally, the Ambiguous galaxies have the most complicated kinematic structures.  They likely contain some combination of less energetic outflows, inflows, rotation, and possibly dual AGNs, although with the longslit data alone we cannot distinguish these individual contributions to the emission line profiles.

When we consider only the galaxies with unambiguous classifications, we find that $77^{+10}_{-12}\%$ of SDSS galaxies at $z>0.1$ with double-peaked narrow AGN emission lines are explained by outflows and $23^{+12}_{-10}\%$ are explained by rotation, where the error bars present the 95\% binomial confidence intervals. These results are consistent, within the confidence intervals, with the results of \cite{NE16.1} for their sample of 71 SDSS galaxies at $z<0.1$ with double-peaked narrow AGN emission lines.  Other studies, which do not use the same kinematic classifications, also find that the majority of double-peaked emission lines are produced by NLR gas kinematics (e.g., \citealt{SH11.1,FU12.1}).  We conclude that most double-peaked narrow AGN emission lines are produced by outflows.

\subsection{Eight Dual AGN Candidates}
\label{dual}

Seven of the galaxies with double-peaked AGN emission lines have companion galaxies that are classified as Seyferts, and one has a companion that is classified as composite. Although the Seyfert-like line flux ratios in each galaxy pair suggests the presence of dual AGNs, there may be a single ionizing source (a single AGN) producing the emission features in both galaxies (e.g., \citealt{MO92.1}).  As a result, these eight merging galaxy systems are candidate dual AGNs. Follow-up high resolution X-ray or radio observations are necessary to resolve whether two AGNs are present and whether the systems indeed host dual AGNs.

Of these eight dual AGN candidates, four (SDSS J0952+2552, SDSS J1245+3723, SDSS J1248$-$0257, and SDSS J1301$-$0058) have a companion galaxy located within the $1\farcs5$ radius SDSS fiber.  These systems are kinematically classified as outflow (SDSS J0952+2552) or ambiguous (SDSS J1245+3723, SDSS J1248$-$0257, and SDSS J1301$-$0058).  However, because both galaxies are located within the SDSS fiber, the motion of dual AGNs could contribute to the double-peaked emission lines in these systems.  The case of dual AGNs where one AGN is driving an outflow and the other AGN is relatively faint could be classified as an outflow, while dual AGNs combined with rotation, inflows, weak outflows, and other kinematic disturbances could be classified as ambiguous \citep{NE16.1}.  There is no clean kinematic classification category for dual AGNs, since the kinematics alone are insufficient to confirm dual AGNs.

For the other four dual AGN candidates, the galaxy pair separations are too large for dual AGNs (where there is one central AGN in each galaxy) to contribute to the double peaks in the SDSS spectra.  

Here, we comment on each dual AGN candidate individually and compare to other analyses published in the literature. We note that these published analyses sometimes define dual AGNs as those that are responsible for the double-peaked emission in the SDSS spectrum, whereas our definition of dual AGNs includes AGN pairs with separations $< 30$ kpc, regardless of whether they contribute to the double peaks. \\

{\bf SDSS J0952+2552} (candidate dual AGNs).  The two stellar bulges seen in the {\it HST}/F160W image are also seen in near-infrared laser guide star adaptive optics $H$-band imaging with the NIRC2 near-infrared camera on the Keck II telescope \citep{RO11.1} and in $H$-band imaging with the OH-Suppressing Infrared Imaging Spectrograph (OSIRIS) on the Keck II telescope \citep{FU12.1}. While previous studies have classified this system as confirmed dual AGNs \citep{MC11.1,FU12.1,MC15.1}, we classify it as candidate dual AGNs because of the need for X-ray or radio confirmation of two AGNs with such a small separation ($1\farcs00$).  \\

{\bf SDSS J1157+0816} (candidate dual AGNs). The galaxy with the double-peaked AGN emission lines has two companions visible in the SDSS imaging: one to the northeast, and one to the southwest.  These three sources are also seen in NIRC2 $K'$-band imaging \citep{FU12.1}.  The galaxy with the double-peaked AGN emission lines and the southwest companion have similar redshifts (Table~\ref{tbl-comp}), but the northeast companion's continuum was too faint in our longslit spectrum to extract a redshift.  We cannot confirm whether it is associated with the other two galaxies, although in the SDSS image all three galaxies appear to be related (Figure~\ref{fig:dualoffset}).  \cite{MC15.1} classified the \oiiiw peaks in SDSS J1157+0816 as unresolved structure with a range of possible explanations, including dual AGNs, NLR kinematics, and small-scale outflows. \\

{\bf SDSS J1245+3723} (candidate dual AGNs). The longslit spectra show continuous emission connecting the two galaxies, and the SDSS image also shows that the companion galaxy has an elongated morphology (Figure~\ref{fig:dualoffset}).  This suggests that the two galaxies may already be interacting. No other studies have published follow-up observations for this object, so it has no previous classifications. \\ 

{\bf SDSS J1248$-$0257} (candidate dual AGNs). NIRC2 $H$-band and $K'$-band imaging of this system reveal the companion galaxy \citep{RO11.1,FU12.1}.  \cite{FU12.1} classified this system as having an extended NLR. \\

{\bf SDSS J1301$-$0058} (candidate dual AGNs). A companion galaxy is seen in {\it HST}/ACS/F550M imaging \citep{FU12.1}, and the companion's elongated tidal tail suggests that the two galaxies are already interacting. \cite{FU12.1} classified this system as having an unresolved NLR. \\

{\bf SDSS J1323+0308} (candidate dual AGNs). The SDSS image shows tidal debris in the system, and the longslit spectrum shows ionized gas forming a bridge between the two galaxies (Figure~\ref{fig:dualoffset}). No other studies have published follow-up observations for this object, so it has no previous classifications. \\

{\bf SDSS J1541+2036} (candidate dual AGNs).  The companion galaxy is seen in NIRC2 $H$-band and $K'$-band imaging \citep{RO11.1,FU12.1}. \cite{FU12.1} classified this system as having an unresolved NLR. \\

{\bf SDSS J1610+1308} (candidate dual AGNs).  The galaxy with the double-peaked AGN emission lines has a companion that is found in NIRC2 $H$-band and $K'$-band imaging \citep{RO11.1,FU12.1}, and a stream of ionized gas connects the galaxy with the double-peaked AGN emission lines to the companion (Figure~\ref{fig:dualoffset}).  \cite{MC15.1} classified the double peaks in \oiiiw as being produced by outflows, and this complicated system may include both outflows and dual AGNs. \\

\subsection{Double-peaked AGN Emission Lines Are Preferentially Associated with Mergers}

Many previous studies have explored the relative AGN fractions in merging and isolated galaxies, as a means of constraining the importance of different AGN triggering mechanisms (e.g., \citealt{EL11.1,SI11.1,KO12.1,LI12.2,SA14.1,FU18.1}).  Here, we also examine the fraction of AGNs in galaxy mergers, but specifically for AGNs that exhibit double-peaked emission lines in their spectra.

Of the eight merging galaxy systems that we find, three of them have both merging galaxies resolved in SDSS imaging (SDSS J1157+0816, SDSS J1245+3723, SDSS J1323+0308).  Consequently, we find that at least $3.2^{+3.0}_{-1.7}\%$ (3/95) of SDSS galaxies with double-peaked AGN emission lines are in galaxy mergers (defined by $| \Delta v | < 500$ km s$^{-1}$ and $\Delta x <30$ kpc) that are resolvable in SDSS imaging.  This is a lower limit, given that the necessary spectroscopic redshifts do no exist for all of the galaxies that have another galaxy located within 30 kpc.  Our lower limit is $\sim2$ times higher than the $1.7^{+0.2}_{-0.2}\%$ of all SDSS AGNs that are in galaxy mergers, defined with the same $| \Delta v |$ and $\Delta x$ cutoffs, and with the same range of galaxy stellar masses (the Kolmogorov-Smirnov probability is $72\%$ that the masses were derived from the same distribution; \citealt{EL08.1}).

The actual fraction of galaxies with double-peaked AGN emission lines that are associated with galaxy mergers is likely much higher.  Studies using $0\farcs1$ resolution imaging have found that $29^{+5}_{-4}\%$ of SDSS galaxies with double-peaked AGN emission lines have another galaxy located within the $3^{\prime\prime}$ SDSS fiber \citep{FU12.1}.

Next, we examine whether our eight merging galaxy systems are in major or minor mergers.  If we define major mergers as mass ratios less than 4:1 and minor mergers as mass ratios greater than 4:1, then we find six major mergers and two minor mergers.  Since the SDSS photometric catalog has a limiting $r$-band magnitude of 22.0, we cannot detect the faintest companion galaxies in our sample.  Consequently, we are underestimating the number of minor mergers.  

A previous study of galaxy pairs in SDSS, which used a similar pair definition ($| \Delta v | < 350$ km s$^{-1}$ and $\Delta x <35$ kpc), explored the relative numbers of major and minor merger galaxy pairs that are detectable in SDSS \citep{LA12.1}.  Using the same mass ratio cutoff as we use between major and minor mergers (4:1), they find that major merger galaxy pairs are twice as common as minor merger galaxy pairs in SDSS.  While this is not indicative of the true mass ratio trend for galaxy mergers (where the minor merger rate is $\sim3$ times larger than the major merger rate; \citealt{LO11.1}), it accounts for bias in the SDSS selection and offers a more accurate comparison for our sample.  For the subset of our galaxy pairs where both galaxies are resolved in SDSS imaging, our results (two major mergers and one minor merger) are consistent with those of the general galaxy pair population in SDSS \citep{LA12.1}.

In six of the dual AGN candidates the more luminous AGN is in the more massive stellar bulge, and in two of the dual AGN candidates the more luminous AGN is in the less massive stellar bulge.  This fits with phenomenological models of dual AGNs that find that $60-70\%$ of dual AGNs have the more luminous AGN in the more massive stellar bulge \citep{YU11.1} and simulations that find that the distinction of more luminous AGN switches between stellar bulges during the course of the galaxy merger \citep{VA12.1}.

\subsection{The More Massive Black Hole Typically Accretes with a Higher Eddington Ratio}

Of the two AGNs in each dual AGN candidate system, we determine which AGN is accreting at a higher rate using the Eddington ratio $f_{Edd} \equiv L_{bol}/L_{Edd}$.   The bolometric luminosity $L_{bol}$ is determined from the \oiiiw luminosity (\citealt{HE04.3}; with a scatter of 0.38 dex), while the Eddington luminosity is $L_{Edd} = 4 \pi c G M_{BH} m_p / \sigma_T$, where $M_{BH}$ is the black hole mass, $m_p$ is the mass of a proton, and $\sigma_T$ is the Thomson scattering cross section.  

If the more luminous stellar bulge in the merging system is ``1" and the less luminous stellar bulge is ``2", then $L_{bol,1}/L_{bol,2}=(f_{Edd,1}/f_{Edd,2}) (M_{BH,1}/M_{BH,2})$.  Assuming that the black hole mass traces the host stellar bulge luminosity \citep{MA03.5,MC04.1,GR07.1}, then $M_{BH,1}/M_{BH,2}=L_{*,1}/L_{*,2}$ and $f_{Edd,1}/f_{Edd,2}=(L_{*,2}/L_{*,1})(L_{bol,1}/L_{bol,2})$.  Using this approach, we estimate $f_{Edd,1}/f_{Edd,2}$ for each dual AGN candidate (Table~\ref{tbl-prop}).

We find that three of the dual AGN candidates have $f_{Edd,1}/f_{Edd,2}<1$, and five have $f_{Edd,1}/f_{Edd,2}>1$.  This implies that the more massive supermassive black hole in a dual AGN system typically accretes with a higher Eddington ratio than the less massive supermassive black holes in a dual AGN system.  Simulations of galaxy mergers find that the less massive black hole accretes with a higher Eddington ratio until the less massive black hole's host galaxy is stripped of its gas \citep{VA12.1,ST16.1}, which implies that in the five systems with $f_{Edd,1}/f_{Edd,2}>1$ (SDSS J1245+3723, SDSS J1301-0058, SDSS J1323+0308, SDSS J1541+2036, SDSS J1610+1308), the less massive galaxy in the merger may have depleted its gas reservoir.  Follow-up observations of the gas contents of the galaxies would confirm whether this is the case. 

\section{Conclusions}
\label{conclusions}

We obtained optical longslit observations of 95 galaxies at $z>0.1$ with double-peaked narrow AGN emission lines in their SDSS spectra.  The data were taken at Lick Observatory, Palomar Observatory, MMT Observatory, Gemini Observatory, and Keck Observatory.  Using the longslit data, we kinematically classified each galaxy as rotation dominated, outflow, or ambiguous.  We also searched the SDSS images (as well as {\it HST} and NIRC2 images, where available) for possible companions to these galaxies, and we analyzed the longslit spectra to determine the redshifts and optical line flux ratios of the galaxies and their companions.

Our main results are summarized below.

\vspace{.1in}

1.  Most double-peaked narrow AGN emission lines are produced by outflows.  Using the kinematics of the \oiiiw emission lines in our longslit spectra, we classify the 95 double-peaked systems as follows: 55\% outflows, 15\% rotation dominated, and 30\% ambiguous.  Although dual AGNs could contribute to the observed emission line profiles, we cannot identify dual AGNs from the gas kinematics alone.

2. We found eight galaxies that have companion galaxies with line-of-sight velocity separations $| \Delta v | < 500$ km s$^{-1}$ and projected physical separations $\Delta x <30$ kpc.  Based on the line flux ratios of the optical emission lines detected in each companion galaxy, all eight systems are dual AGN candidates.  Three of these merging galaxy systems are resolved in SDSS imaging, while five required high spatial resolution near-infrared imaging (with {\it HST} or NIRC2) to resolve the companion galaxies.

3. Active galaxies with double-peaked narrow emission lines in their spectra are found in galaxy mergers at least twice as often as active galaxies in general are found in galaxy mergers.  At least $3\%$ of SDSS galaxies with double-peaked narrow AGN emission lines are in galaxy mergers where both galaxies are resolved by SDSS imaging. 

4.  In five of the dual AGN candidates, the more massive of the two supermassive black holes accretes gas with the higher Eddington ratio.  Since simulations of galaxy mergers show that the less massive black hole accretes with a higher Eddington ratio until the less massive black hole's host galaxy loses its gas, this suggests that in these five systems the less massive galaxy may have already been tidally stripped of its gas. 

\vspace{.1in}

The eight candidate dual AGNs presented here could be confirmed as dual AGNs via the detection of two AGN emission sources, using high spatial resolution X-ray or radio observations.  Radio observations would also reveal any radio jets associated with the AGN outflows. 

\acknowledgements J.M.C. thanks Emma Hogan for guidance during the Gemini data reduction.  We also thank the anonymous referee for comments that have improved the clarity of this paper.

The Gemini Observatory is operated by the Association of Universities for Research in Astronomy, Inc., under a cooperative agreement with the NSF on behalf of the Gemini partnership: the National Science Foundation (United States), the National Research Council (Canada), CONICYT (Chile), the Australian Research Council (Australia), Minist\'{e}rio da Ci\^{e}ncia, Tecnologia e Inova\c{c}\~{a}o (Brazil) and Ministerio de Ciencia, Tecnolog\'{i}a e Innovaci\'{o}n Productiva (Argentina). GMOS data were obtained under programs GN-2010A-Q-91, GS-2010A-Q-67, GN-2010B-Q-79, and GS-2010B-Q-32.
The W. M. Keck Observatory is operated as a scientific partnership among the California Institute of Technology, the University of California, and the National Aeronautics and Space Administration. The Observatory was made possible by the generous financial support of the W. M. Keck Foundation.  We also wish to recognize and acknowledge the highly significant cultural role and reverence that the summit of Mauna Kea has always had within the indigenous Hawaiian community. It is a privilege to be given the opportunity to conduct observations from this mountain.

{\it Facilities:} \facility{Gemini:North (GMOS)}, \facility{Gemini:South (GMOS)},  \facility{Keck:I (LRIS)}, \facility{Keck:II (DEIMOS)}

\bibliographystyle{apj}

\end{document}